\documentclass[10pt]{article}

\usepackage{amsmath,amssymb,graphicx}
\def\be{\begin{equation}}
\def\ee{\end{equation}}
\def\d{{\rm d}}
\def\e{{\rm e}}

\begin{document}

\begin{center}
{\Large\bf Non-orthogonally transitive $G_2$ spike solution}
\vspace{.3in}
\\{\bf Woei Chet Lim}
\\Department of Mathematics, University of Waikato, Private Bag 3105, Hamilton 3240, New Zealand
\\Email: wclim@waikato.ac.nz
\vspace{.1in}
\vspace{0.2in}

\end{center}

\begin{abstract}
  
We generalize the orthogonally transitive (OT) $G_2$ spike solution to the non-OT $G_2$ case.
This is achieved by applying Geroch's transformation on a Kasner seed. The new solution contains two more parameters than the OT $G_2$ spike solution.
Unlike the OT $G_2$ spike solution, the new solution always resolves its spike. 
\end{abstract}

[PACS: 98.80.Jk, 04.20.-q, 04.20.Jb]

\section{Introduction}

According to general relativity, in the asymptotic regime near spacelike singularities, a spacetime would oscillate between Kasner states.
The BKL conjectures~\cite{art:LK63,art:BKL1970,art:BKL1982} hold except where and when spikes occur~\cite{art:BergerMoncrief1993,art:GarfinkleWeaver2003}. 
Spikes are a recurring inhomogeneous phenomenon in which the fabric of spacetime temporarily develops a spiky structure as the spacetime oscillates between Kasner states.
See the introduction section of~\cite{art:HeinzleUgglaLim2012} for a comprehensive background.

Previously in \cite{art:Lim2008} the orthogonally transitive (OT) $G_2$ spike solution, which is important in describing the recurring spike oscillation,
was generated by applying the Rendall-Weaver transformation~\cite{art:RendallWeaver2001} on a Kasner seed solution.
The solution is unsatisfactory, however, in that it contains permanent spikes,
and there is a debate whether permanent spike are actually unresolved spike transitions in the oscillatory regime or are really permanent.
In other words, would the yet undiscovered non-OT $G_2$ spike solution contain permanent spikes?
The proponents for permanent spikes argue that the spatial derivative terms of a permanent spike are negligible, and hence the spike stays permanent~\cite{art:Garfinkle2007}.
The opponents base their argument on numerical evidence that the permanent spike is mapped by an $R_1$ frame transition to a regime where the spatial derivative terms are not neglibigle, 
which allows the spike to resolve~\cite{art:HeinzleUgglaLim2012}. 
To settle the debate, we need to find the non-OT $G_2$ spike solution.
It was found that Geroch's transformation~\cite{art:Geroch1971,art:Geroch1972} would generate the desired solution, which always resolves its spike.
The next section describes the generation process.

\section{Generating the solution}

For our purpose, we express a metric $g_{ab}$ using the Iwasawa frame~\cite{art:HeinzleUgglaRohr2009}, as follows. Indicies $0,1,2,3$ corresponds to coordinates $\tau,x,y,z$.
Assume zero vorticity (zero shift). The metric components in terms of $b$'s and $n$'s are given by
\begin{align}
        g_{00} &= -N^2
\\
        g_{11} &= e^{-2b_1},\quad g_{12} = e^{-2b_1} n_1,\quad g_{13} = e^{-2b_1} n_2
\\
        g_{22} &= e^{-2b_2} + e^{-2b_1} n_1^2,\quad g_{23} =  e^{-2b_1} n_1 n_2 + e^{-2b_2} n_3
\\
        g_{33} &= e^{-2b_3} + e^{-2b_1} n_2^2 + e^{-2b_2} n_3^2.
\end{align}
One advantage of the Iwasawasa frame is that the determinant of the metric is given by 
\be
	\det g_{ab} = -N^2 e^{-2b_1-2b_2-2b_3}.
\ee

A pedagogical starting point is the Kasner solution with the following parametrization:
\be
        b_1 = \frac14(w^2-1)\tau,\quad
        b_2 = \frac12(w+1)\tau,\quad
        b_3 = -\frac12(w-1)\tau,\quad
        N^2 = \e^{-2b_1-2b_2-2b_3} = \e^{-\frac12(w^2+3)\tau},
\ee
and $n_1=n_2=n_3=0$.
We shall use a linear combination of all three Killing vector fields (KVFs)
\be
	a_1 \partial_x + a_2 \partial_y + a_3 \partial_z.
\ee
as the KVF in Geroch's transformation, so that the transformation generates the most general metric possible from the given seed.

\subsection{Change of coordinates}

To simplify the KVF before applying Geroch's transformation, make the coordinate change
\be
	x = X + n_{10} Y + n_{20} Z,\quad y = Y + n_{30} Z ,\quad z = Z
\ee
where $n_{10}$, $n_{20}$, $n_{30}$ are constants. 
Then the metric parameters $b_1$, $b_2$, $b_3$ and $N$ are unchanged but $n_1=n_{10}$, $n_2=n_{20}$, $n_3=n_{30}$ are now constants instead of zero.
The KVF becomes
\be
	(a_3 (n_{10} n_{30} - n_{20}) - a_2 n_{10}  + a_1) \partial_X + (a_2 - a_3 n_{30}) \partial_Y + a_3 \partial_Z.
\ee
We cannot set the $Z$ component to zero, but we can set the $X$ and $Y$ components to zero, leading to
\be
	n_{30} = \frac{a_2}{a_3},\quad n_{10} = \frac{a_1}{a_3}.
\ee
Without loss of generality, we set $a_3=1$, and so $n_{30} = a_2$ and $n_{10} = a_1$. $n_{20}$ remains free. We will see later that it can be used to eliminate any $y$-dependence.

To make transparent the effect of Geroch's transformation on the $b$'s (see (\ref{nonOT_spike})--(\ref{nonOT_spike_b3}) below), it is best to adapt the KVF to $\partial_x$.
So we make another coordinate change to swap $X$ and $Z$:
\be
	X = \tilde{z},\quad Y = \tilde{y},\quad Z = \tilde{x},
\ee
which in effect introduces frame rotations to the Kasner solution. The Kasner solution now has
\begin{align}
\label{Kasner_rotated}
	N^2 &= \e^{-\frac12(w^2+3)\tau}
\\
	\e^{-2b_1} &= \e^{(w-1)\tau} + n_{20}^2 \e^{-\frac12(w^2-1)\tau} + n_{30}^2 \e^{-(w+1)\tau}
\\
	\e^{-2b_2} &= \frac{\mathcal{A}^2}{\e^{-2b_1}}
\\
	\e^{-2b_3} &= \e^{-\frac12(w^2+3)\tau} \mathcal{A}^{-2}
\\
	n_1 &= \frac{n_{30} \e^{-(w-1)\tau} + n_{10} n_{20} \e^{-\frac12(w^2-1)\tau}}{\e^{-2b_1}}
\\
	n_2 &= \frac{n_{20} \e^{-\frac12(w^2-1)\tau}}{\e^{-2b_1}}
\\
\label{Kasner_rotated_n3}
	n_3 &= \e^{-\frac12(w^2-1)\tau}\mathcal{A}^{-2}\left[n_{30}(n_{10}n_{30}-n_{20})\e^{-(w+1)\tau}+n_{10}\e^{(w-1)\tau} \right],
\intertext{where}
\label{area}
        \mathcal{A}^2 &= (n_{10} n_{30} - n_{20})^2 \e^{-\frac12(w+1)^2\tau} + n_{10}^2 \e^{-\frac12(w-1)^2\tau} + \e^{-2\tau}.
\end{align}

Effectively, we are applying Geroch's transformation to the seed solution (\ref{Kasner_rotated})--(\ref{Kasner_rotated_n3}), using the KVF $\partial_{\tilde{x}}$.
We shall now drop the tilde from the coordinates.

\subsection{Applying Geroch's transformation}

Applying Geroch's transformation using a KVF $\xi_a$ involves the following steps.
First compute
\be
	\lambda=\xi^a \xi_a
\ee
and integrate the equation
\be
	\nabla_a \omega =\varepsilon_{abcd}\xi ^b\nabla^c \xi^d
\ee
for the general solution for $\omega$. $\omega$ is determined up to an additive constant $\omega_0$.
In our case we get
\be
	\lambda = \e^{-2b_1} = \e^{(w-1)\tau} + \e^{-\frac12(w^2-1)\tau} n_{20}^2 + \e^{-(w+1)\tau} n_{30}^2,\quad
	\omega = 2w n_{30} z - K y + \omega_0,
\ee
where the constant $K$ is given by
\be
	K = \frac12 (w-1)(w+3) n_{20} - 2 w n_{10} n_{30}.
\ee
We could absorb $\omega_0$ by a translation in the $z$ direction if $w n_{30} \neq 0$, but we shall keep $\omega_0$ for the case $w n_{30} = 0$.

The next step involves finding a particular solution for $\alpha_a$ and $\beta_a$:
\begin{align}
	\nabla_{[a}\alpha_{b]} &=\frac{1}{2}\varepsilon_{abcd} \nabla^c \xi^d,\quad \xi^a \alpha_a =\omega,
\\
	\nabla_{[a}\beta_{b]} &=2\lambda \nabla_a \xi_b + \omega \varepsilon_{abcd}  \nabla^c \xi^d,\quad \xi^a \beta_a =\omega^2 + \lambda^2 - 1.
\end{align}
Without loss of generality, we choose $\theta=\frac{\pi}{2}$ in Geroch's transformation, so $\alpha_a$ is not needed in $\eta_a$ below.
We assume that $\beta_a$ has zero $\tau$-component. Its other components are
\begin{align}
	\beta_1 &= \omega^2 + \lambda^2 -1
\\
        \beta_2 &= n_{10} n_{20}^3 \e^{-(w^2-1)\tau} + \left[ 2 \frac{w-1}{w+1} n_{10} n_{20} n_{30}^2 + \frac{4}{w+1} n_{20}^2 n_{30} \right] \e^{-\frac12(w+1)^2 \tau}
\notag\\
        &\quad
                + 2 \frac{w+1}{w-1} n_{10} n_{20} \e^{-\frac12(w-1)^2\tau} + (w+1) n_{30} \e^{-2\tau} + n_{30}^3 \e^{-2(w+1)\tau} + F_2(y,z)
\\
        \beta_3 &= n_{20}^3 \e^{-(w^2-1)\tau} + 2 n_{20} n_{30}^2 \frac{w-1}{w+1} \e^{-\frac12(w+1)^2 \tau} + 2 n_{20} \frac{w+1}{w-1} \e^{-\frac12(w-1)^2\tau} + F_3(y,z)
\end{align}
where $F_2(y,z)$ and $F_3(y,z)$ satisfy the constraint equation
\be
        - \partial_z F_2 + \partial_y F_3 + 2(w-1)\omega = 0.
\ee
For our purpose, we want $F_3$ to be as simple as possible, so we choose
\be
	F_3 = 0,\quad F_2 = \int 2(w-1)\omega \d z = 2w(w-1) n_{30} z^2 - 2 (w-1) K y z +2(w-1)\omega_0 z.
\ee

The last step constructs the new metric. Define $\tilde{\lambda}$ and $\eta_a$ as
\begin{align}
	\frac{\lambda}{\tilde{\lambda}} &= (\cos\theta-\omega\sin\theta)^2 +\lambda^2 \sin^2\theta,
\\
	\eta_a &=\tilde{\lambda}^{-1} \xi_a +2 \alpha_a \cos\theta\sin\theta-\beta_a \sin^2\theta.
\end{align}
The new metric is given by
\be
	\tilde{g}_{ab}=\frac{\lambda}{\tilde{\lambda}}(g_{ab}-\lambda^{-1} \xi_a\xi_b)+\tilde{\lambda} \eta_a \eta_b.
\ee
In our case $\tilde{g}_{ab}$ is given by the metric parameters
\begin{align}
\label{nonOT_spike}
        \tilde{N}^2 &= N^2 (\omega^2+\lambda^2)
\\
        \e^{-2\tilde{b}_1} &= \frac{\e^{-2b_1}}{\omega^2+\lambda^2}
\\
        \e^{-2\tilde{b}_2} &= \e^{-2b_2} (\omega^2+\lambda^2)
\\
\label{nonOT_spike_b3}
        \e^{-2\tilde{b}_3} &= \e^{-2b_3} (\omega^2+\lambda^2)
\\
        \tilde{n}_1 &= -2w(w-1) n_{30} z^2 + 2 (w-1) K y z - 2(w-1)\omega_0 z + \frac{\omega^2}{\lambda}(n_{30} \e^{-(w+1)\tau} + n_{10} n_{20} \e^{-\frac12(w^2-1)\tau})
\notag\\
        &\quad  -\left[ n_{30} w \e^{-2\tau} + \frac{w+3}{w-1} n_{10} n_{20} \e^{-\frac12(w-1)^2\tau} + \frac{w-3}{w+1} n_{20} n_{30} (n_{10} n_{30} - n_{20}) \e^{-\frac12(w+1)^2\tau} \right]
\\
        \tilde{n}_2 &= n_{20} \e^{-\frac12(w^2-1)\tau} \left[ - \frac{w+3}{w-1} \e^{(w-1)\tau} - n_{30}^2 \frac{w-3}{w+1} \e^{-(w+1)\tau} + \frac{\omega^2}{\lambda} \right]
\\
        \tilde{n}_3 &= \mathcal{A}^{-2} \left[ n_{10} \e^{-\frac12(w-1)^2\tau} + n_{30} (n_{10} n_{30} - n_{20}) \e^{-\frac12(w+1)^2\tau} \right],
\label{nonOT_spike_n3}
\end{align}
and $\mathcal{A}$, given by (\ref{area}), is the area density~\cite{art:vEUW2002} of the $G_2$ orbits. 
Note that the $w=\pm1$ cases would have to be computed separately, which we shall leave to future work.
The new solution admits two commuting KVFs:
\be
        \partial_x,\quad [- (w-1)K^2y^2+2(w-1)K\omega_0 y] \partial_x + 2wn_{30}\partial_y + K \partial_z.
\ee
Their $G_2$ action is non-OT, unless $n_{10}=n_{20}=0$.
The solution is also the first non-OT Abelian $G_2$ explicit solution found.

In the next section we shall focus on the case where $K=0$, or equivalently, where
\be
\label{n20choice}
	n_{20} = \frac{4w}{(w-1)(w+3)} n_{10} n_{30},
\ee
which turns off the $R_2$ frame transition (which is shown  to be asymptotically suppressed in~\cite{art:HeinzleUgglaRohr2009}), and eliminates the $y$-dependence.
Setting (\ref{n20choice}) in the rotated Kasner solution (\ref{Kasner_rotated})--(\ref{Kasner_rotated_n3}) also turns off the $R_2$ frame transition there, giving
the explicit solution that describes the double frame transition $\mathcal{T}_{R_3 R_1}$ in~\cite{art:HeinzleUgglaRohr2009}.
The mixed frame/curvature transition $\mathcal{T}_{N_1 R_1}$ in~\cite{art:HeinzleUgglaRohr2009} is described by the metric $\tilde{g}_{ab}$
with $n_{20} = n_{30} = 0$. Both the double frame transition and the mixed frame/curvature transition are encountered in the exceptional Bianchi type VI$_{-1/9}^*$
cosmologies~\cite{art:Hewittetal2003}.

Setting $n_{10}=n_{20}=0$ yields the OT $G_2$ spike solution in~\cite{art:Lim2008}. To adapt the solutions in~\cite{art:Lim2008} to the Iwasawa frame here, let
\be
	b_1 = - \frac12 (P(\tau,z)-\tau), \quad b_2 = \frac12 (P(\tau,z)+\tau), \quad b_3 = - \frac14 (\lambda(\tau,z)+\tau), \quad n_1 = -Q(\tau,z), \quad n_2=n_3=0,
\ee
where $x$-dependence in~\cite{art:Lim2008} becomes $z$-dependence here, and set $w$ to $-w$, $\lambda_2=\ln 16$, $Q_0 = 1$, $Q_2 = 0$ there, and set $n_{30}=1$, $\omega_0=0$ here.
As pointed out in~\cite{art:Limetal2009} and~\cite{art:Binietal2009}, the factor 4 in Equation (34) of~\cite{art:Lim2008} should not be there.

\section{The dynamics of the solution}

To describe the dynamics of the non-OT spike solution, we shall plot the state space orbit
projected onto the Hubble-normalized $(\Sigma_+,\Sigma_-)$ plane, as done in~\cite{art:Lim2008}.
The formulas are
\begin{align}
	\Sigma_+ &= -1 + \frac14 \mathcal{N}^{-1} \partial_\tau (\mathcal{A}^2)
\\
	\Sigma_- &= \frac{1}{2\sqrt{3}} \mathcal{N}^{-1} \partial_\tau(\tilde{b}_2 - \tilde{b}_1)
\\
	\mathcal{N} &= \frac16\left[ \frac{\partial_\tau (\lambda^2)}{\omega^2+\lambda^2} + \partial_\tau \ln (N^2) \right]
\end{align}
\cite{art:HeinzleUgglaRohr2009} uses a different orientation, where their $(\Sigma_+,\Sigma_-)$ are given by
\begin{align}
	\Sigma_+ &= -\frac12(\Sigma_+ + \sqrt{3}\Sigma_-)
\\
	\Sigma_- &= -\frac12(\sqrt{3}\Sigma_+ - \Sigma_-)
\end{align}

The non-OT spike solution (with $K=0$, $\omega_0=0$) goes from a Kasner state with $2 < w < 3$, through a few intermediate Kasner states, and arrives at the final Kasner state with $w < -1$.
The transitions are composed of spike transitions and $R_1$ frame transitions. 
The non-OT spike solution always resolves its spike, unlike the OT spike solution with $|w|<1$, which has a permanent spike.

For a typical Kasner source with $2 < w < 3$, there are six non-OT spike solutions, some of which are equivalent, that start there. For example,
non-OT spike solutions with $|w| = \tfrac13,\ 2,\ 5$ all start at $w_\text{source} = \frac73$. From there, however, there are two extreme alternative spike orbits. The first
alternative is to form a ``permanent" spike, followed by an $R_1$ transition, and lastly to resolve the spike. This was described in~\cite{art:HeinzleUgglaLim2012} as the joint spike transition.
This alternative is more commonly encountered (assuming that permanent spikes are more commonly encountered than no-spike at the end of a Kasner era).
The second alternative is to undergo an $R_1$ transition first, followed by a transient spike transition, and finish with another $R_1$ transition.
By varying $n_1$ and $n_3$, one can get orbits that are close to one extreme alternative or the other, or some indistinct mix.

\begin{figure}
  \begin{center}
    \resizebox{\textwidth}{!}{\includegraphics{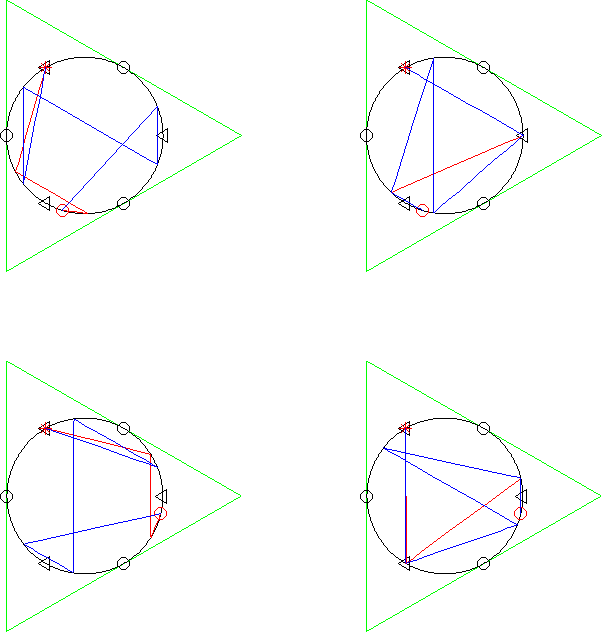}}
    \caption{Alternative spike orbits for $w=5$. Top row is the orientation used in~\cite{art:Lim2008}, bottom row is the orientation used in~\cite{art:HeinzleUgglaRohr2009}. 
Left column is the first alternative orbit, right column is the second alternative. 
Spike orbits ($z=0$) are in red, faraway orbits ($z=10^{12}$) in blue. 
Left column is generated with $n_{10}=10^{-3}$, $n_{30}=1$, right column with $n_{10}=10^9$, $n_{30}=10^{-9}$. 
A red circle marks the start of the orbits, a red star marks the end.}
    \label{fig:w5alternatives}
\end{center}
\end{figure}

The sequence of $w$-value of the Kasner states for the spike orbit is given below.
For non-OT spike solution with $|w|>3$, the first and second alternatives are
\begin{align}
        &\frac{3|w|-1}{1+|w|},\ \frac{5+|w|}{1+|w|},\ 2+|w|,\ 2-|w|
\\
        &\frac{3|w|-1}{1+|w|},\ \frac{3|w|+1}{|w|-1},\ \frac{|w|-5}{|w|-1},\ 2-|w|
\end{align}
For $1<|w|<3$, the first and second alternatives are
\begin{align}
        &\frac{5+|w|}{1+|w|},\ \frac{3|w|-1}{1+|w|},\ \frac{3|w|+1}{|w|-1}, \frac{5-|w|}{1-|w|}
\\
        &\frac{5+|w|}{1+|w|},\ 2+|w|,\ 2-|w|, \frac{5-|w|}{1-|w|}
\end{align}
For $|w|<1$, the first and second alternatives are
\begin{align}
        &\frac{5+|w|}{1+|w|},\ \frac{3|w|-1}{1+|w|},\ \frac{3|w|+1}{|w|-1}, \frac{5-|w|}{1-|w|}
\\
        &\frac{5+|w|}{1+|w|},\ 2+|w|,\ 2-|w|, \frac{5-|w|}{1-|w|}
\end{align}
For $|w|<1$, the first and second alternatives are
\begin{align}
        &2+|w|,\ 2-|w|,\ \frac{5-|w|}{1-|w|},\ \frac{3|w|+1}{|w|-1}
\\
        &2+|w|,\ \frac{5+|w|}{1+|w|},\ \frac{3|w|-1}{1+|w|},\ \frac{3|w|+1}{|w|-1}
\end{align}
For example, for $|w| = \tfrac13,\ 2,\ 5$, the first alternative is $\frac73,\ \frac53,\ 7,\ -3$ and the second alternative is $\frac73,\ 4,\ 0,\ -3$.
See Figure~\ref{fig:w5alternatives}.

\section{Summary}

In this paper, we went through the steps of generating the non-OT $G_2$ spike solution, and illustrated its state space orbits for the case $K=0$, which show two extreme alternative orbits.
More importantly, the non-OT $G_2$ spike solution always resolves its spikes, in contrast to its OT $G_2$ special case which produces an unresolved permanent spike for some parameter values.
The non-OT $G_2$ spike solution shows that, in the oscillatory regime near spacelike singularities, unresolved permanent spikes are artefacts of restricting oneself to the OT $G_2$ case, 
and that spikes are resolved in the more general non-OT $G_2$ case. Therefore spikes are expected to recur in the oscillatory regime rather than to become permanent spikes.
We also obtained explicit solutions describing the double frame transition and the mixed frame/curvature transition in~\cite{art:HeinzleUgglaRohr2009}.
We leave the further analysis of the non-OT $G_2$ spike solution to future work.

\section*{Acknowledgment}

Part of this work was carried out at the Max Planck Institute for Gravitational Physics (Albert Einstein Institute) and Dalhousie University.
I would like to thank Claes Uggla and Alan Coley for useful discussions.
The symbolic computation software {\tt MAPLE} and numerical software {\tt MATLAB} are essential to the work.

\bibliography{}

\end{document}